\documentclass[letterpaper]{article} 
\usepackage{aaai23}  
\usepackage{times}  
\usepackage{helvet}  
\usepackage{courier}  
\usepackage[hyphens]{url}  
\usepackage{graphicx} 
\usepackage{natbib}  
\usepackage{caption} 
\usepackage{algorithm}
\usepackage{algorithmic}

\newcommand{\red}[2]{{{#1}}}
\usepackage{url}

\usepackage{newfloat}
\usepackage{listings}
\DeclareCaptionStyle{ruled}{labelfont=normalfont,labelsep=colon,strut=off} 
\floatstyle{ruled}
\newfloat{listing}{tb}{lst}{}
\floatname{listing}{Listing}
\usepackage{cite}
\usepackage{amsmath,amssymb,amsfonts}
\usepackage{algorithmic}
\usepackage{graphicx}
\usepackage{textcomp}
\usepackage{xcolor}
\usepackage{booktabs}
\usepackage{diagbox}
\usepackage{multirow}
\usepackage{subfigure}
\usepackage{caption}
\usepackage{enumitem}

\newcommand{\tb}{\noindent\textbf}
\newcommand{\ie}{\emph{i.e.}}
\newcommand{\eg}{\emph{e.g.}}

\title{DACOM: Learning Delay-Aware Communication for Multi-Agent\\ Reinforcement Learning}
\author {
    Tingting Yuan\textsuperscript{\rm 1}\thanks{Corresponding author},
    Hwei-Ming Chung\textsuperscript{\rm 2,3},
    Jie Yuan\textsuperscript{\rm 4},
    Xiaoming Fu\textsuperscript{\rm 1}
}
\affiliations {
    \textsuperscript{\rm 1} University of G{\"o}ttingen,
    \textsuperscript{\rm 2} University of Oslo,  \\
    \textsuperscript{\rm 3} NOOT Tech. Co., Ltd., 
    \textsuperscript{\rm 4} Beijing University of Posts and Telecommunications\\
    tingting.yuan@cs.uni-goettingen.de,
    hmc@noot.ai,
    yuanjie@bupt.edu.cn,
    fu@cs.uni-goettingen.de
}

\usepackage{bibentry}

\begin{document}

\maketitle

\begin{abstract}
Communication is supposed to improve multi-agent collaboration and overall performance in cooperative Multi-agent reinforcement learning (MARL). However, such improvements are prevalently limited in practice since most existing communication schemes ignore communication overheads (\eg, communication delays). In this paper, we demonstrate that ignoring communication delays has detrimental effects on collaborations, especially in delay-sensitive tasks such as autonomous driving. To mitigate this impact, we design a delay-aware multi-agent communication model (DACOM) to adapt communication to delays. Specifically, DACOM introduces a component, TimeNet, that is responsible for adjusting the waiting time of an agent to receive messages from other agents such that the uncertainty associated with delay can be addressed. Our experiments reveal that DACOM has a non-negligible performance improvement over other mechanisms by making a better trade-off between the benefits of communication and the costs of waiting for messages.
\end{abstract}

\section{Introduction}

Reinforcement learning (RL) \cite{arulkumaran2017deep} enables an agent to interact with an environment, and the agent improves its policy iteratively by learning from observations to achieve a given goal.
RL, with a single agent to decide the behavior of all entities, faces various challenges, such as scalability \cite{yan2021acc} and privacy issues \cite{yuan2022pp}.
To this end, the extension from single-agent RL to multi-agent RL (MARL) \cite{hernandez2019survey} is favorable.
MARL \cite{hernandez2019survey} has been widely used in various tasks, such as real-time resource allocation \cite{yuan2020dynamic}, smart grid control \cite{chung2020distributed}, and autonomous driving \cite{kiran2021deep}.

A major challenge in deploying MARL to solve cooperative tasks is that the partial observation of agents hinders collaboration due to the uncertainty and incompleteness in observing and the non-stationary behavior \cite{sukhbaatar2016learning, zhang2021multi}.
For example, blind spots in autonomous driving can be deadly and lead to collisions between a vehicle and another vehicle, bicyclist, or pedestrian.
Agent communication for information exchange is expected to be a desirable alternative to overcome this challenge by improving cooperation between agents.
However, the improvement is prevalently limited in practice due to the limitations and overheads of communication channels (\eg, limited bandwidth and communication delay) \cite{zhu2022survey}.

Most of the literature has been devoted to designing an effective communication model for multi-agent cooperation to improve the gains of communication, but the impacts of \textit{dynamic communication channels} and \textit{communication overheads} (\eg, delays) have been largely ignored.
Firstly, bandwidth constraints are considered in existing works \cite{mao2020learning,kim2019learning} by scheduling communication, which learns to select agents to share information under limited and constant bandwidth.
However, in practical scenarios, the communication abilities (\eg, bitrate) of agents are heterogeneous and dynamic due to the variability of communication channels.
For instance, in wireless networks \cite{agiwal2016next}, agents' bitrates are related to the allocated bandwidth and the signal-to-interference-plus-noise ratio (SINR), which vary according to the distance from agents to the access points.
Secondly, the communication delay can interfere with the cooperation between agents by introducing delays in action-making \cite{chen2021delay} and uncertainty on the arrival time of information.
Previous work \cite{kim2019learning} prevents endless waiting by setting a predefined and constant bound for the waiting time, but it may restrain potential cooperation if it is set too short and conversely may cause meaningless waiting.
Therefore, such a constant timer is inflexible and cannot be adapted to the dynamics in the communication networks.


To improve agent cooperation, each agent should determine not only whether it is worth waiting but also how long it can wait, taking into account the state of the communication channel.
Motivated by this, this paper proposes a delay-aware multi-agent communication model (DACOM) to realize efficient communication by scheduling the waiting time of communication that is adaptive to dynamic network states.
DACOM aims to achieve a good trade-off between the benefits of communication and \red{the costs of delayed response due to waiting for messages.
}{}To this end, we introduce TimeNet, a novel component in DACOM, to determine whether it is worth waiting and how long it will wait based on the network states and local observation.


To verify DACOM's effectiveness, we conduct extensive experiments in different environments: particle games, traffic control in autonomous driving, and StarCraftII. 
We modify the original environments to support delay-aware actions.
Furthermore, we apply DACOM and baselines in different communication channels with delays ranging from 10\% to 90\% of step intervals.
Our experiments show that DACOM outperforms other baseline mechanisms.
Moreover, we confirm that ignoring communication delays in communication scheduling results in fewer rewards, which is more evident in delay-sensitive tasks whose performance is worse than non-communication MARL.

\noindent \textbf{Contributions}. 
To the best of our knowledge, this paper is the first work that studies the effect of communication delay on MARL for agent cooperation.
We provide theoretical analysis and experimental verification about it.
Additionally, we demonstrate that ignoring communication delay results in performance degradation in delay-sensitive tasks, and the performance is even worse than in non-communication mechanisms in high-latency networks.
Next, we propose a novel model for delay-aware communication that can improve agent cooperation by scheduling waiting time to adapt to network states. 

\section{Background}

\textbf{Reinforcement Learning (RL)}. RL can be formulated based on Markov Decision Process (MDP) as a tuple $<s, a, r, P>$ with state $s$, action $a$, reward $r$, and transition probability function $P$. The transition probability function $P: s \times a \mapsto s$ maps states and actions to a probability distribution over the next states. The goal of RL is to learn a policy by maximizing the expected discounted return $R_t = \mathbb{E}[\sum_{k=0}^{\infty} \gamma^k r_{t+k}]$, where $\gamma$ is a discount factor for future rewards to dampen their effect.

\noindent  \textbf{Deep Q-Networks (DQN)}. 
DQN \cite{mnih2015human} is a combination of RL and a deep neural network that learns the action-value function $Q(s,a) = \mathbb{E}[R|s_t=s, a_t=a]$.
During the training phase, DQN is trained by minimizing the loss $\mathcal{L}(\theta) = \mathbb{E}[(y'-Q(s,a))^2]$, where $y' = r + \gamma \max_{a'} Q'(s',a')$ for the following state and action. Here, $Q'$ is the target network, which is designed to stabilize the training, and its parameters are copied from $Q$ periodically.
During execution, the agent selects the action maximum action-value $\arg\max_a Q(s,a)$.

\noindent \textbf{Policy Gradient (PG)}. 
PG \cite{sutton1999policy} is different from DQN, whose parameters $\theta$ of the policy $\pi$ are directly adjusted by maximizing the objective $J(\theta) = \mathbb{E}[R]$ along the direction of policy gradient $\nabla_{\theta}J(\theta) = \mathbb{E}[\nabla_{\theta} \log\pi_{\theta}(a|s) Q(s,a)]$.
It can be further extended to Deterministic Policy Gradient (DPG) \cite{2014DPG} with deterministic policies $\mu_{\theta}$. The policy network is updated with $\nabla_{\theta}J(\theta) = \mathbb{E}[\nabla_{\theta}\mu_{\theta}(a|s) \nabla_a Q(s,a)|_{a=\mu_{\theta}(s)}]$.

\noindent \textbf{Deep Deterministic Policy Gradient (DDPG)}.
DDPG \cite{lillicrap2015continuous} is an actor-critic algorithm with two deep neural networks to approximate the deterministic policy and the action-value function, respectively.
The policy network, also called the actor, infers actions according to states $a=\mu(s; \theta^{\mu})$.
The Q-network, called the critic, approximates the action-value function $Q(s,a;\theta^Q)$.
The actor is updated with the gradient $\nabla_{\theta}J(\theta^{\mu})$ as DPG, and the critic is updated with loss function $\mathcal{L}(\theta^Q) = \mathbb{E}[(r +\gamma Q'(s',a')-Q(s,a))^2]$.

\noindent \textbf{Partially Observable Markov}.
An environment is partially observable if it cannot be fully and continuously observed. 
MARL in a partially observable environment can be formulated as a decentralized partially observable Markov decision process (Dec-POMDPs) \cite{oliehoek2016concise}, which is an extension of MDP for multiple agents who may have partial observations. 
A Dec-POMDP can be defined as a tuple $<N, S, A, R, P, O, \Omega, \gamma >$,
which includes the number of agents $N$, the space of global states $S$, the set of actions $A =\{a_1, ..., a_N\}$, the set of individual rewards $R = \{r_1, ..., r_N\}$ with $r_i: o_i\times a_i \rightarrow \mathbb{R}$, transition function $P: S \times A \rightarrow S$, a set of local agents' observations $O = \{o_1, ..., o_N\}$ following observation function $\Omega: S \rightarrow O$, and the discount factor $\gamma \in [0, 1]$. Each agent aims to maximize its own total expected return $\mathbb{E}[\sum_{k=0}^{\infty} \gamma^k r_{i, t+k}]$. In fully cooperative games, agents share common rewards, for example, team-average rewards $R = \frac{1}{N} \sum_{i=1}^N r_i$.

\red{
\noindent \textbf{Delay-aware Modeling}.
Delay-aware MDP and Delay-aware Markov Game have been proposed in \cite{chen2021delay} and \cite{chen2020delay}, respectively.
In \cite{chen2021delay} and \cite{chen2020delay}, delays are considered in states and action delays are measured in terms of the number of steps.
However, inter-agent communication or scheduling communication when modeling is not considered.
Moreover, the case of delayed actions executed within the current step is not included.
This kind of case is not rare, as with the support of high-speed communications (\eg, 5G and 6G), it is increasingly possible to complete communications and then execute an action within one step.
}{}

\noindent \textbf{Multi-Agent Deep Deterministic Policy Gradient (MADDPG)}.
Applying DDPG to the multi-agent environment is challenging because the input dimension of the actor-network increases when more agents are presented.
To this end, the authors in \cite{lowe2017multi} proposed a framework containing a centralized critic and distributed actors.
In MADDPG, each agent can make actions with local observations, and the actors' update needs the global states to reach the maximum reward.
The critic is then deployed in a centralized manner to acquire the states from all agents.
The critic creates the action-value function with the states from all agents to guide actors' updates.
The update procedure of MADDPG is extended from DDPG as $\nabla_{\theta_{i}}J(\theta_{i}) = \mathbb{E}[\nabla_{\theta_{i}}\mu_{\theta}(a_{i}|o_{i}) \nabla_a Q(O, A)|_{a_i=\mu_{\theta}(o_i)}]$, where the subscript $i$ indicates agent $i$.
The critic is trained by minimizing the loss $\mathcal{L}(\theta) = \mathbb{E}[(y'-Q(S,A))^2]$, where $y' = r + \gamma Q'(S',A')$.

\section{Methods}

In this section, we will explain the design of DACOM in detail.
The aim of DACOM is to improve the efficiency and robustness of inter-agent communication based on practical communication channels, where communication costs cannot be ignored.
The main idea of DACOM is to adaptively set the waiting time of messages from other agents, given the states of the communication channels.

\subsection{Communication Channels}
In a partially observable environment, agents may not completely perceive the environment that obstructs overall learning and action-making.
To this end, the agents utilize communication to share local observations to improve overall performance.
In practical scenarios, agents are connected by an underlying communication network, such as a wireless or wired network.
The agents' communication abilities (e.g., bitrate) are heterogeneous and dynamic due to the variability in communication networks.
In the following, we analyze the communication capabilities of agents using wireless communication as an example.

In wireless networks, the communication ability is related to many factors, such as signal strength and distance from the source \cite{agiwal2016next}.
As defined by \cite{goldsmith2005wireless}, the bitrates (in bps) from device $i$ to device $j$ is denoted as $x_{i,j} = B_{i,j}\log_2(1+\eta_{i,j})$, where $B$ is the allocated bandwidth (Hz) and $\eta$ is SINR.
The SINR can be defined as $\eta_{i,j} =\frac{\rho_{i,j}}{10^{\varphi_{i,j}/10}(\sigma^2+I^2)}$, where $\rho$ is the transmission power (w); $\sigma ^{2}$ is the additive white Gaussian noise power at the receiver; $I^2$ is the interference; and $\varphi$ is an approximation of distance-dependent path loss.
The delay between two devices can be defined as $l_{i,j} = \frac{m}{x_{i,j}}$, where $m$ is the size of messages in bits.
Note that devices here can be either agents or communication infrastructures, such as base stations (in wireless networks) and roadside units (in vehicular networks).
According to this definition, the communication delay and bitrate are relative to distance, transmission power, and allocated bandwidth, which are dynamic and vary between agents.
\red{Therefore, the communication scheduling of existing works, such as \cite{mao2020learning,kim2019learning}, is done under a static communication state (\eg, a constant bandwidth) so that the methods in these works are not universally applicable.}{}

\subsection{Delay-Aware Modeling}
Previous works model MARL with communication as Dec-POMDP.
Specifically, at each time step, agents make actions based on partial observations and messages from other agents, and the actions are assumed to be executed immediately without any delay.
However, introducing agent communication for cooperation raises the communication delay, which leads to action delay \red{(\ie, the time between observing and taking action).}{}
The action delay changes the interaction between the agents and the environment, especially for delay-sensitive tasks.
\red{
This is because action delays lead to more reaction time of agents, which then reduce rewards (\eg, longer arrival times due to slow acceleration) and even bring penalties (\eg, collisions in an emergency).
}{}

\noindent \tb{DACOM-MDP}.
Given the above analysis,
delays are critical for agent cooperation and cannot be ignored.
Thus, we define the delay-aware partially observable MDP (DACOM-MDP) extended from Dec-POMDP.
DACOM-MDP is described as a tuple $<N, S, A, D, R, P, O, \Omega, \gamma>$. 
We add action delay $D=\{\tilde{d}_i\}_{i = 1, ..., N}$ as a new element and modify reward function as $R: O\times A \times D \rightarrow \mathbb{R}$.
Then, we augment state space and observation space to include network states as $S=\{s, s^{net}\}$ and $O = \{ o_i, o_i^{net}\}_{i=1, ..., N}$.
Specifically, network states can be an end-to-end communication delay $o^{net}_i=\{l_{i,j}\}_{j=1, ..., N}$ or bitrate $o^{net}_i=\{x_{i,j}\}_{j=1, ..., N}$ between agents.
\red{Given the challenge in observing current network states before the action making and timer setting, it can be estimated by using the recent observations of networks.
For example, the weighted moving average has been widely used for networking state estimation \cite{kumar2020swift}), which can be used to estimate current communication delay.}{}

\tb{Theoretical analysis}.
We use value-based MARL as an example to illustrate the advantages of DACOM-MDP.
In previous work, which ignored communication states and action delays, the agents' objective is modeled as $\hat{Q}(O, A)$.
In contrast, our objective for optimizing communication between agents is to maximize $Q(O,O^{net},A,D)$, which is a delay-aware action-value.
The effect of ignoring communication states and action delays in the modeling is analyzed below.

\noindent \textbf{Proposition 1}
\textit{
Consider an action-value function $\hat{Q}(O,A)$ that is modeled without considering network states and action delay.
We can derive a lower bound for the loss between the delay-aware action-value function $Q^*(O, O^{net}, A,D)$ and $\hat{Q}(O, A)$ as
}
\begin{equation}
\hspace{-1mm}
\mathcal{L}(\hat{Q},Q^*)\geq \sum_{i,j\in N} \underbrace{\mathbb{E}[V_h(m_{i,j})}_{\text{Gains}}] p(l_{i,j} \geq d_i) + \underbrace{\mathbb{E}[V_d(D)]}_{\text{Costs}},
\hspace{-0.5mm}
\label{L_QV_1}
\end{equation}
\textit{where $V_h$ is the gain of messages $m_{i,j}$; $p$ is the probability of the arrival of message $m_{i,j}$ after action-making; and $V_d$ is the cost of delays $D$
}(\textit{Proof.} See Appendix).

\noindent In particular, we extend Proposition 1 to two specific cases: 

\noindent \textbf{Proposition 2}
\textit{
Consider a network delay follows normal distribution $l_{i,j} \sim \mathcal{N}(\lambda, \sigma^2)$, we analyze two cases: distributed communication mode (\ie, exchange of information between agents, denoted as $\hat{Q}_a$) and the centralized communication mode (\ie, a message aggregator located centrally to gather all messages from agents, denoted as $\hat{Q}_b$). We have the lower bound of loss on action-value follows:}
\begin{equation}
\mathcal{L}^*(\hat{Q}_a,Q^*)\propto  \lambda + \xi_n \sigma, \;\;\;
\mathcal{L}^*(\hat{Q}_b,Q^*) \propto  2\lambda + \xi_n \sigma,
\label{eq2_1}
\end{equation}
\textit{where $\xi_n$ is the inverse of the standard normal cumulative distribution function (CDF), with $n\leq N$ agents' communication
}
(\textit{Proof.} See Appendix).

According to Propositions 1 and 2, we can conclude that ignoring communication delays deteriorates the learning performance in multi-agent scenarios, leading to arbitrarily suboptimal policies and further causing non-stationary results.
In Equation \eqref{L_QV_1}, the lower bound of the loss increases with the importance of communication $V_h$, the rate of missing messages $p$, the delay-sensitivity $V_d$, and the action delay $D$.
Given Equation \eqref{eq2_1}, the centralized communication mode has more delays than the decentralized communication mode. 
Therefore, a decentralized communication mode is preferred when designing DACOM.

\subsection{Architecture Design}

\begin{figure}[h!]
\centering
\includegraphics[width=0.47\textwidth] {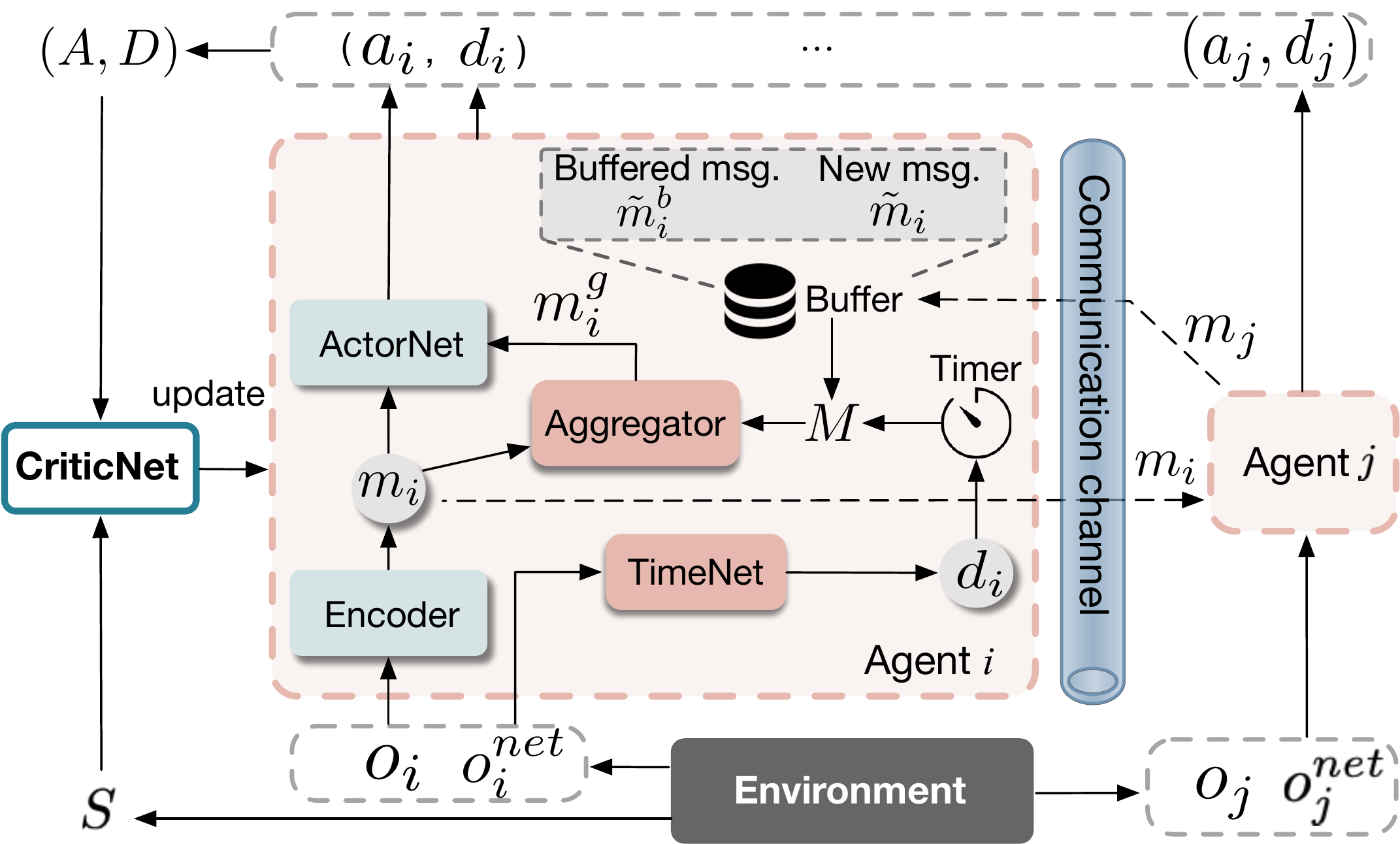}
\caption{DACOM: The Encoder encodes observation $o$ into messages $m$, the TimeNet schedules the waiting time $d$, the buffer caches the latest messages, the aggregator aggregates messages $M$ into $m^g$, and the ActorNet makes actions $a$ based on messages.}
\label{fig:DACOM_arch}
\end{figure}

The proposed DACOM is an adaptively delay-aware agent communication model, where agents learn to schedule waiting time (\ie, how long to wait for communication considering dynamic network states) \red{by introducing a novel component--TimeNet.}{}
DACOM is extended from the actor-critic, which adopts centralized learning and decentralized execution.
As shown in Fig. \ref{fig:DACOM_arch}, each agent consists of three components: 1) an actor with two parts (Encoder and ActorNet); 2) a TimeNet for waiting-time scheduling (\ie, setting timers $d$ for message waiting);
and 3) an attention-based aggregator for message aggregation.
They are parameterized by $\theta^{\mu}$, $\theta^{\tau}$, and $\theta^{g}$, respectively.
The CriticNet, denoted by $\theta^Q$, is a centralized critic network that estimates delay-aware action-value $Q$ for training agents.
Deep neural networks are applied to implement all these components.

In DACOM, at each time step, the Encoder in agent $i$ takes the local observation, $O_i=\{o_i, o_i^{net}\}$, as input, and then encodes messages for agent communication, denoted by $m_i= \mu^{enc}(o_i, o_i^{net}; \theta_i^{enc})$.
At the same time, the TimeNet has the same input as the Encoder, and it sets the waiting thresholds $d_i= \tau(o_i, o_i^{net}; \theta_i^{\tau})$ to timer.
Agent $i$ receives messages from other agents that arrive within time $d_i$.
The new arrival messages in $t$ are denoted by $\tilde{m}_{i,t} = \{m_{j,t}|l_{i,j,t}\leq d_{i,t}\}$, where $l_{i,j}$ is the end-to-end communication delay from agent $j$ to agent $i$.
A buffer is designed to store the latest messages (called buffered messages $\tilde{m}_i^{b}$) given that messages may have high continuity and similarity in some environment \cite{zhang2020succinct}.
In this case, agents can use the buffered messages without waiting for current messages $m_{j, t}$ by setting $d_{i,t}=0$.
Therefore, the available messages is $M_i=\{\tilde{m}_i, \tilde{m}_i^{b}\}$.
There is an attention-based aggregator that aggregates messages $\{m_{i}, M_i\}$, whose output is denoted by $m_i^{g} = g(m_i, M_i; \theta_i^{g})$.
Besides, the ActorNet is introduced for action selection, which takes local message $m_i$ and aggregated message $m_i^{g}$ as input and actions $a_i = \mu^{act}(m_i, m_i^{g}; \theta_i^{\mu})$ as output.
In DACOM, the waiting time $d_i$ can be viewed as a part of actions, and the joint action is $\{A,D\}=\{a_i,d_i\}_{i = 1,...,N}$.
As the rest part (\eg, the time cost by shallow neural networks) can be ignored, the action delay $\tilde{d}_i$ is assumed to approximate communication delay $d_i$.
The action-value estimated by CriticNet is $Q(O,O^{net},A,D)$, which is related to the action delay and network states.

\red{
The contribution of DACOM is the incorporation of $o^{net}$ into learning communications and the active scheduling of waiting times via TimeNet.
The waiting time scheduling is a trade-off between gains of communication and costs of action delays.
On one hand, setting a greater delay tolerance (\ie, higher $d$) can bring more messages, however, it will result in more action delays and eventually reduce the overall gains of communication.
On the other hand, setting a lower $d$ may result in important messages being missing and good cooperation not being achieved.
Intuitively, it is worth waiting a long time when communication can greatly increase the rewards.
In the case of autonomous driving, for example, if there is a car with blind views and the inter-agent communication is able to share the missing information about them to avoid a collision, the gains brought by communication will be obviously high.
On the contrary, if communication cannot bring benefit to agents, a shorter waiting time is preferred.
For example, if the vehicle already has enough views to ensure safe driving, the communication gain will be limited (\eg, a few minutes earlier arrival) or even negative (\eg, collisions due to more response time).
}{}

\subsection{Training}

For training DACOM, an experience reply buffer is required.
The experience replay buffer $\mathcal{H}$ contains tuples $<O,O^{net}, A, D, R, O', O'^{net}>$ that are recorded from agents' experiences, including joint actions are $\{ A, D\}$, rewards $R$, and next observations $O'$ and $O'^{net}$.

\noindent \textbf{Critic}. The joint delay-aware action-value function is formulated as $Q(O,O^{net},A,D; \theta^Q) = \mathbb{E}[\sum_{t=0}^{\infty} \gamma^t r_t]$.
The weights $\theta^Q$ are updated by minimizing the loss $\mathcal{L}(\theta^{Q})= \mathbb{E} [(Q(O,O^{net}, A,D)-y)^2]$ and $ y = R+\gamma Q(O', O'^{net}, A',D')$, 
where $A'$ and $D'$ are given by target actors and target TimeNets with the subsequent observations $O'$ and $O'^{net}$.

\noindent \textbf{Actor}. 
Actors need to learn to maximize the objective $J(\theta^{\mu}, \theta^g, \theta^\tau)$.
An actor, which contains an Encoder and an ActorNet, is denoted by $\mu_i=\{\mu^{enc}_i, \mu^{act}_i\}$ with the parameters $\theta^{\mu}_i =\{\theta_i^{enc}, \theta_i^{act}\}$.
\red{The training of actors uses the deterministic policy gradient, and}{}
the gradient for updating the parameters of agent $i$ can be written as:
\begin{gather}
\nabla_{\theta_i^{\mu}}J(\theta_i^{\mu},\cdot,\cdot ) = \mathbb{E}_{O, O^{net}, D, A_i^-\sim \mathcal{H}} [\nabla_{\theta_i^{\mu}}\mu_i(o_i, o_i^{net}, m_i^g)\notag\\
\nabla_{a_i} Q(O,O^{net},A, D)|_{a_i=\mu_i (o_i, o_i^{net}, m_i^g)}],
\label{theta}
\hspace{-2em}
\end{gather}
where $A_i^{-}$ is the joint action except for agent $i$.

\noindent \textbf{Aggregator}.
By applying the chain rule, the aggregator gradient can be further derived based on $\nabla_{\theta_i^{\mu}}J(\cdot,\cdot,\theta_i^{\mu})$ as:
\begin{align}
\notag
\nabla_{\theta_i^g}J(\cdot, \theta_i^g,\cdot ) 
= \mathbb{E}_{O,O^{net}, D, A_i^-\sim \mathcal{H}} [\nabla_{\theta_i^g} g_i(m_i,M_i) \\
\notag
\nabla_{\theta_i^{\mu}}\mu_i(o_i, o_i^{net}, m_i^g)|_{m_i^g=g(m_i,M_i)}\\
 \nabla_{a_i} Q(O,O^{net},A, D)|_{a_i=\mu_i (o_i, o_i^{net},m_i^g)}].
 \label{g}
\end{align}
The aggregator $g$, Encoder, and ActorNet can be optimized jointly using back-propagation (see Equations \eqref{theta} and \eqref{g}).

\noindent \textbf{TimeNet}. TimeNet aims to select an appropriate waiting time, $d_i$, for each agent to maximize delay-aware action value $Q$. 
\red{The TimeNet can be viewed as a part of the entire actor in DACOM, whose gradient can be represented as:}{}
\begin{align}
\notag
\nabla_{\theta_i^{\tau}}J(\cdot, \cdot, \theta_i^{\tau})
= \mathbb{E}_{O,O^{net}, D_i^-, A^-\sim \mathcal{H}} [\nabla_{\theta_i^{\tau}}\tau_i(o_i, o_i^{net}) \\
\nabla_{d_i} Q(O,O^{net}, A, D)|_{d_i = \tau(o_i, o_i^{net}),a_i=\mu_i}],
\end{align}
where $D_i^{-}$ is the joint timer of the agents except for agent $i$.
When training TimeNet, $d_i$ is given by $\tau(o_i, o_i^{net})$, and $d_i$ results in new $M_i$, $m_i^g$, and finally $a_i$.

\noindent \textbf{Target Networks}.
Finally, the weights of target networks are periodically updated softly with $\theta'_i \leftarrow \xi  \theta_i + (1-\xi) \theta'_i$, where $\xi$ is a coefficient between 0 and 1. This function works with $\theta$, including $\theta_i^\mu$, $\theta_i^g$, $\theta_i^{\tau}$, and $\theta_i^Q$.

\section{Experiments}
\label{exp}
In this section, we evaluate the performance of DACOM in three well-known environments.

\subsection{Environments}
The evaluation environments are multi-agent particle games
\footnote{https://github.com/openai/multiagent-particle-envs.},
autonomous driving
\footnote{https://github.com/eleurent/highway-env.},
and StarCraft Multi-Agent Challenge (SMAC)\footnote{https://github.com/oxwhirl/smac.}.
We modify the original environment to include communication delays, reflect the effects of action delays, and give agents different capabilities in observing.
\begin{figure}[htp!]
\centering
\subfigure[co-navigation] {
 \label{fig:cn}     
\includegraphics[width=0.145\textwidth]{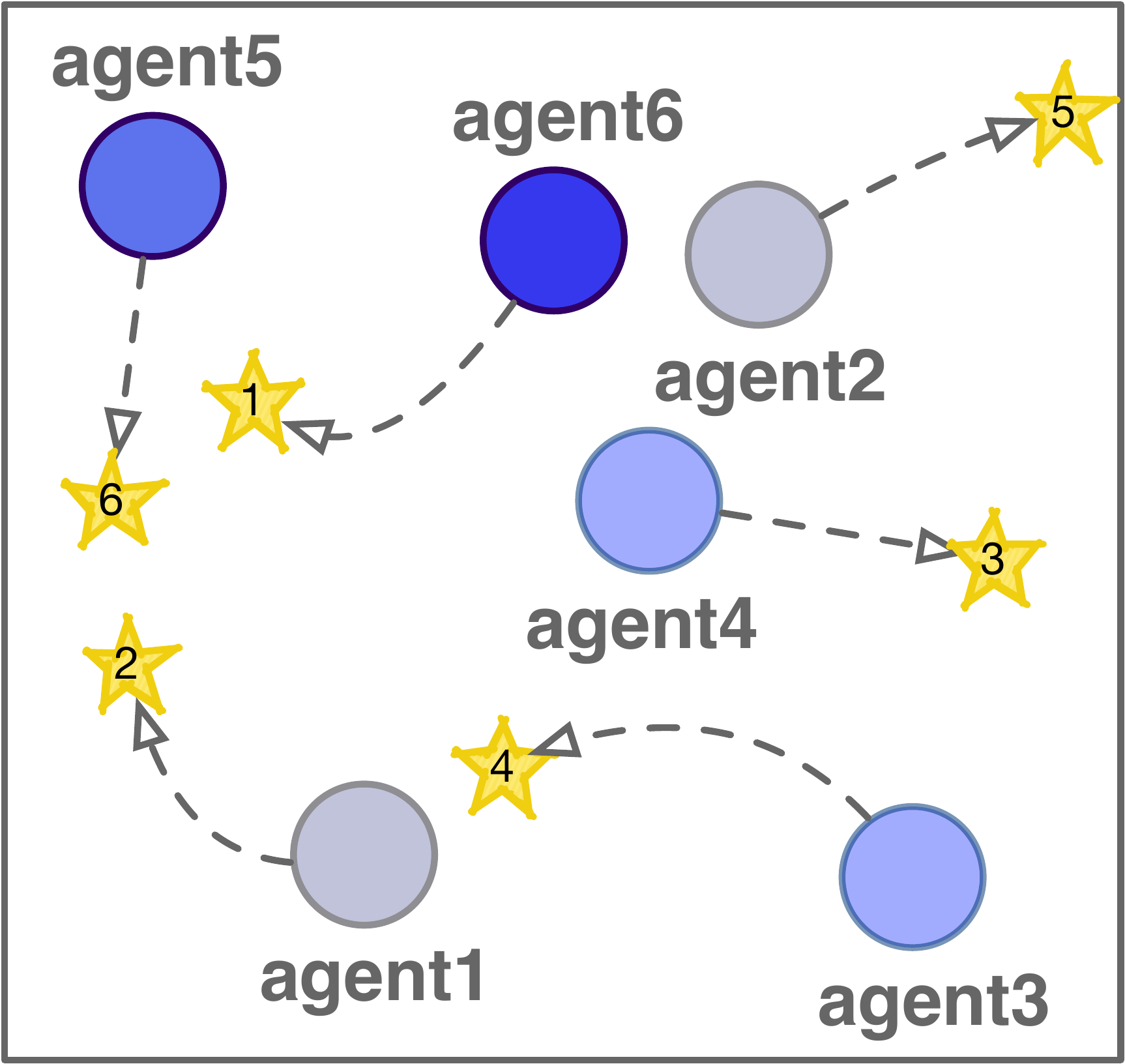}
}
\hspace{-0.5em}
\subfigure[predator-prey] {
 \label{fig:pp}     
\includegraphics[width=0.145\textwidth]{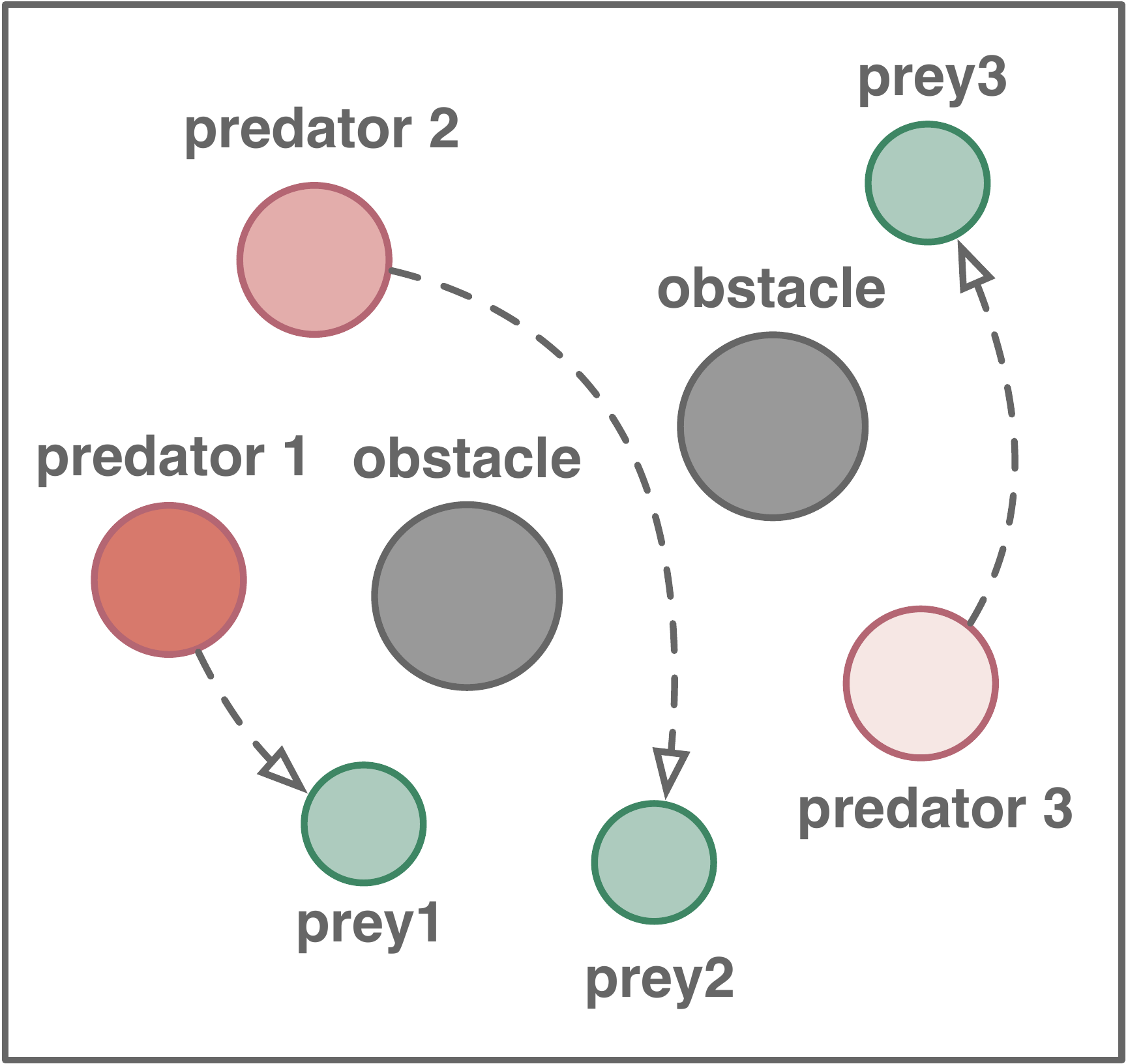}
}
\hspace{-0.5em}
\subfigure[traffic control] {
 \label{fig:tc}     
\includegraphics[width=0.145\textwidth]{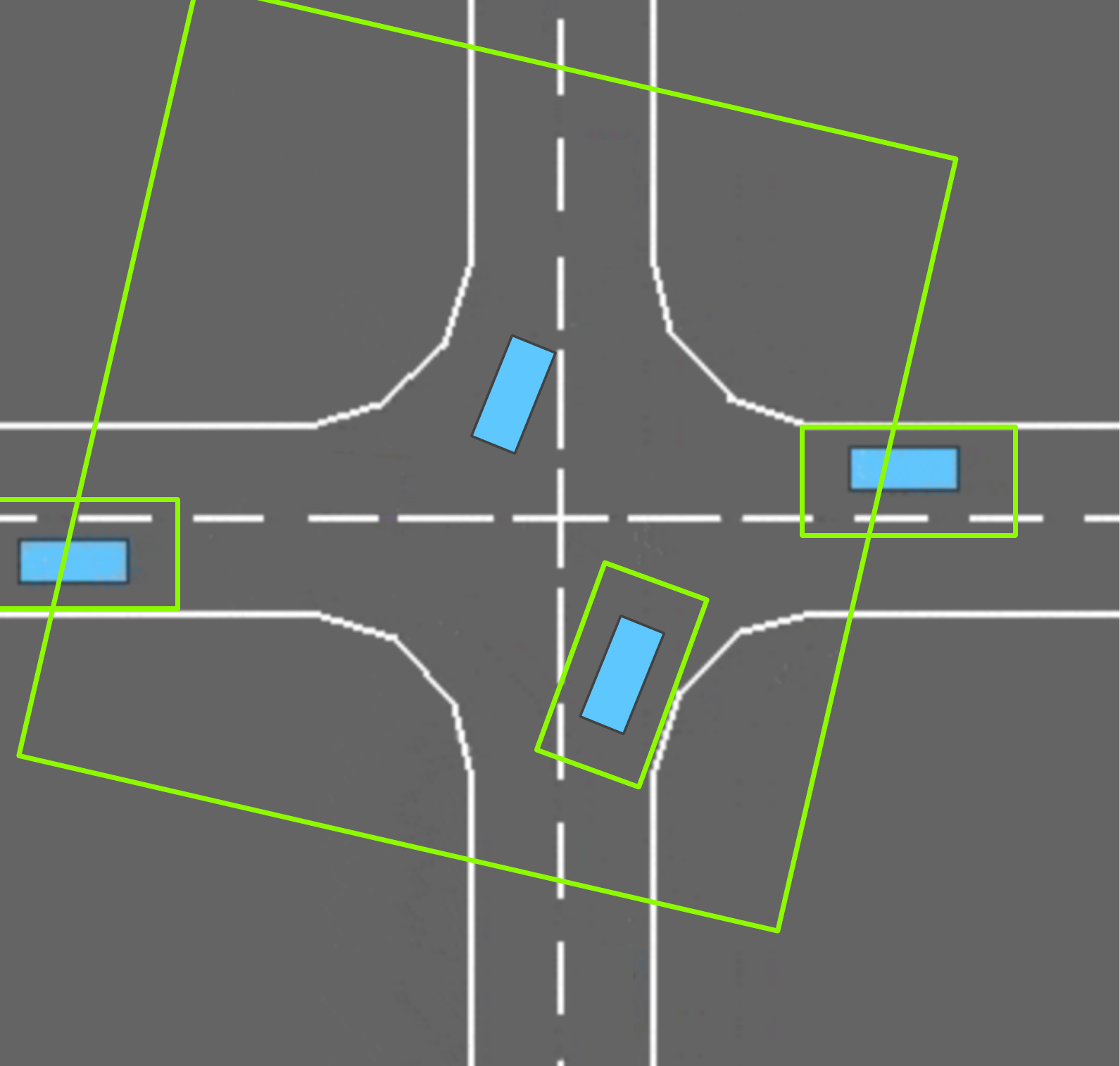}
}
\\
\subfigure[SMAC 3m] {
 \label{fig:sc2-a}     
\includegraphics[width=0.226\textwidth]{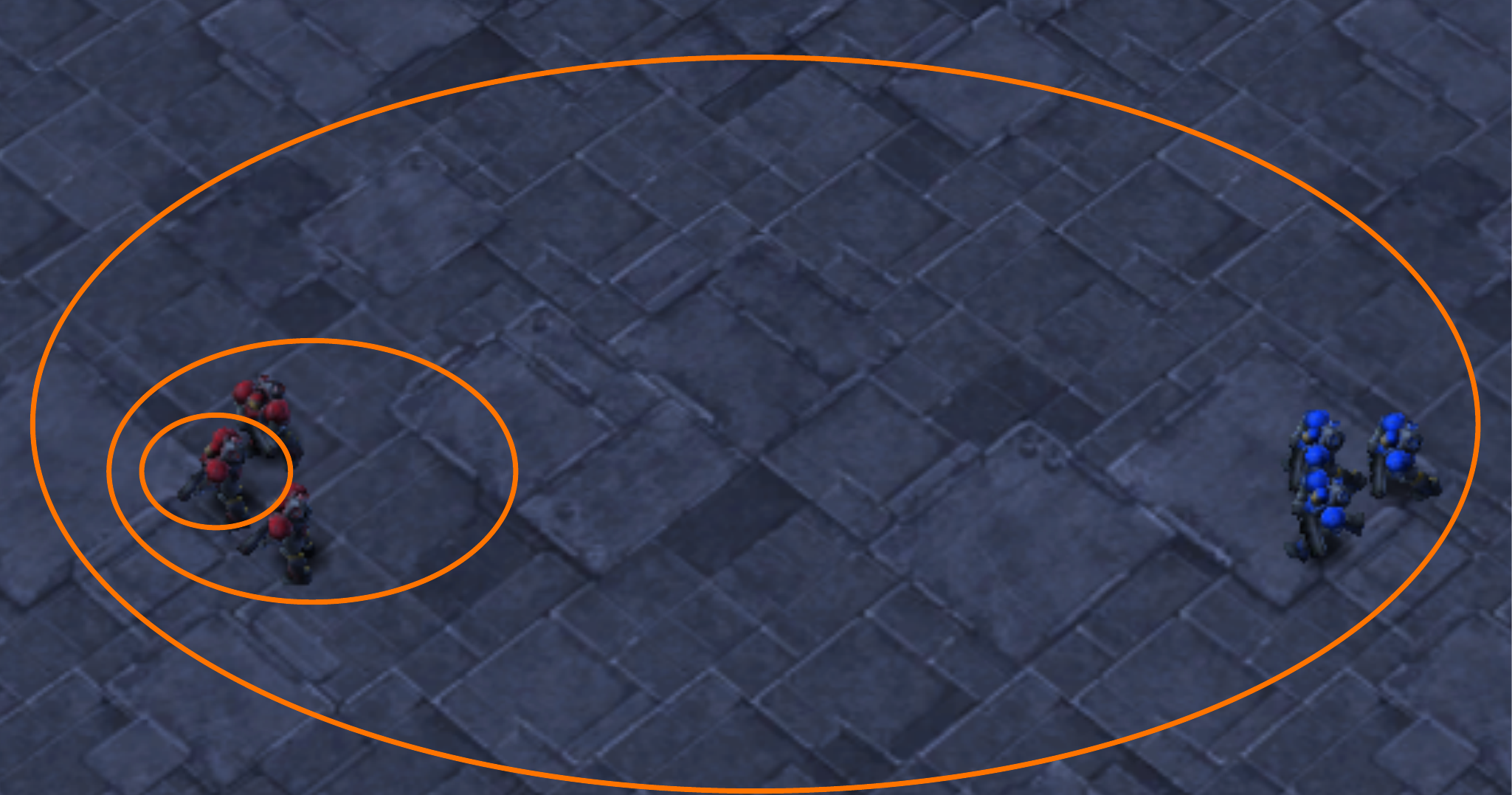}
}
\hspace{-0.5em}
\subfigure[SMAC 8m] {
 \label{fig:sc2-b}     
\includegraphics[width=0.226\textwidth]{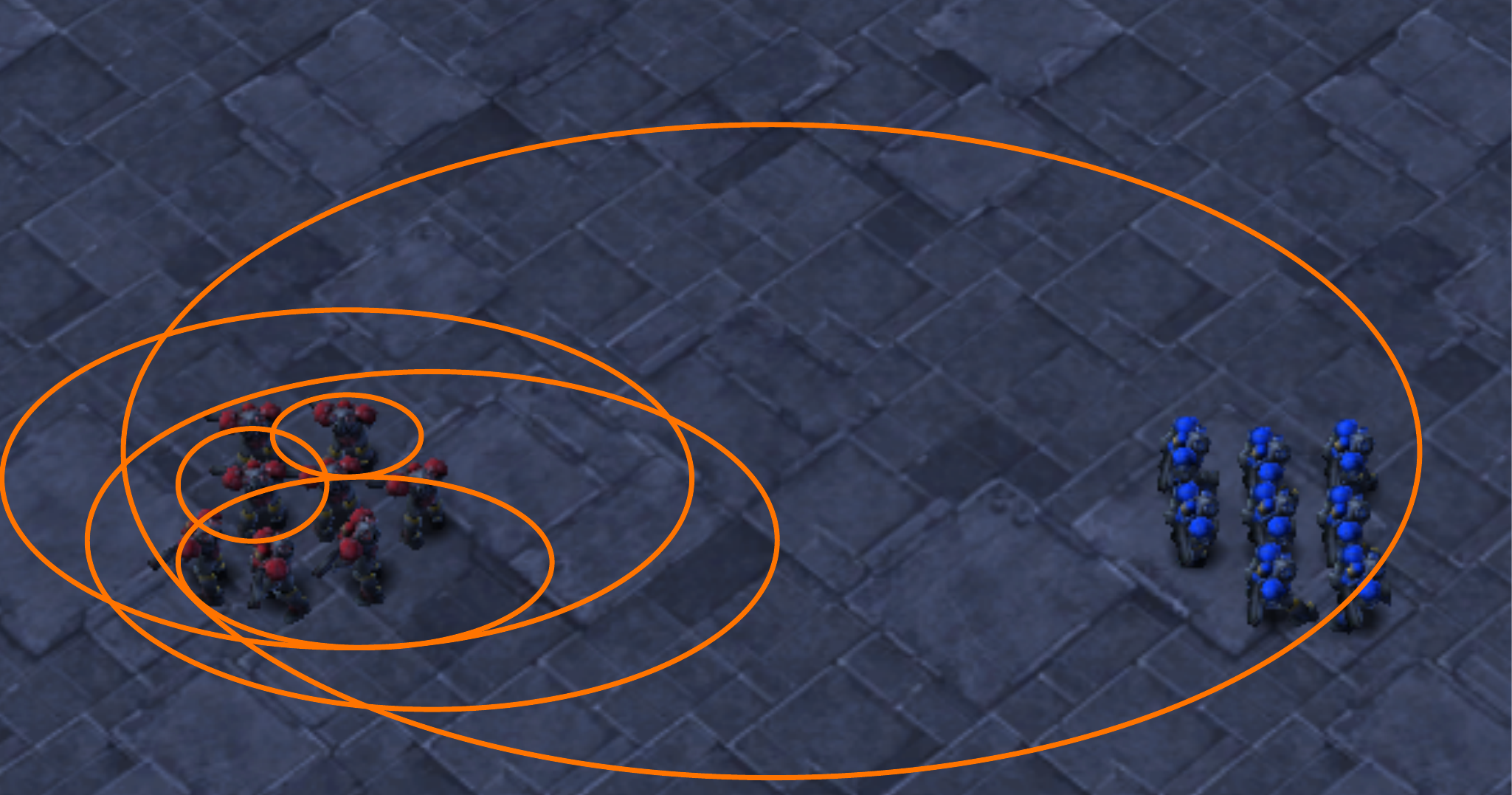}
}
\subfigure[SMAC 2s3z] {
 \label{fig:sc2-c}     
\includegraphics[width=0.226\textwidth]{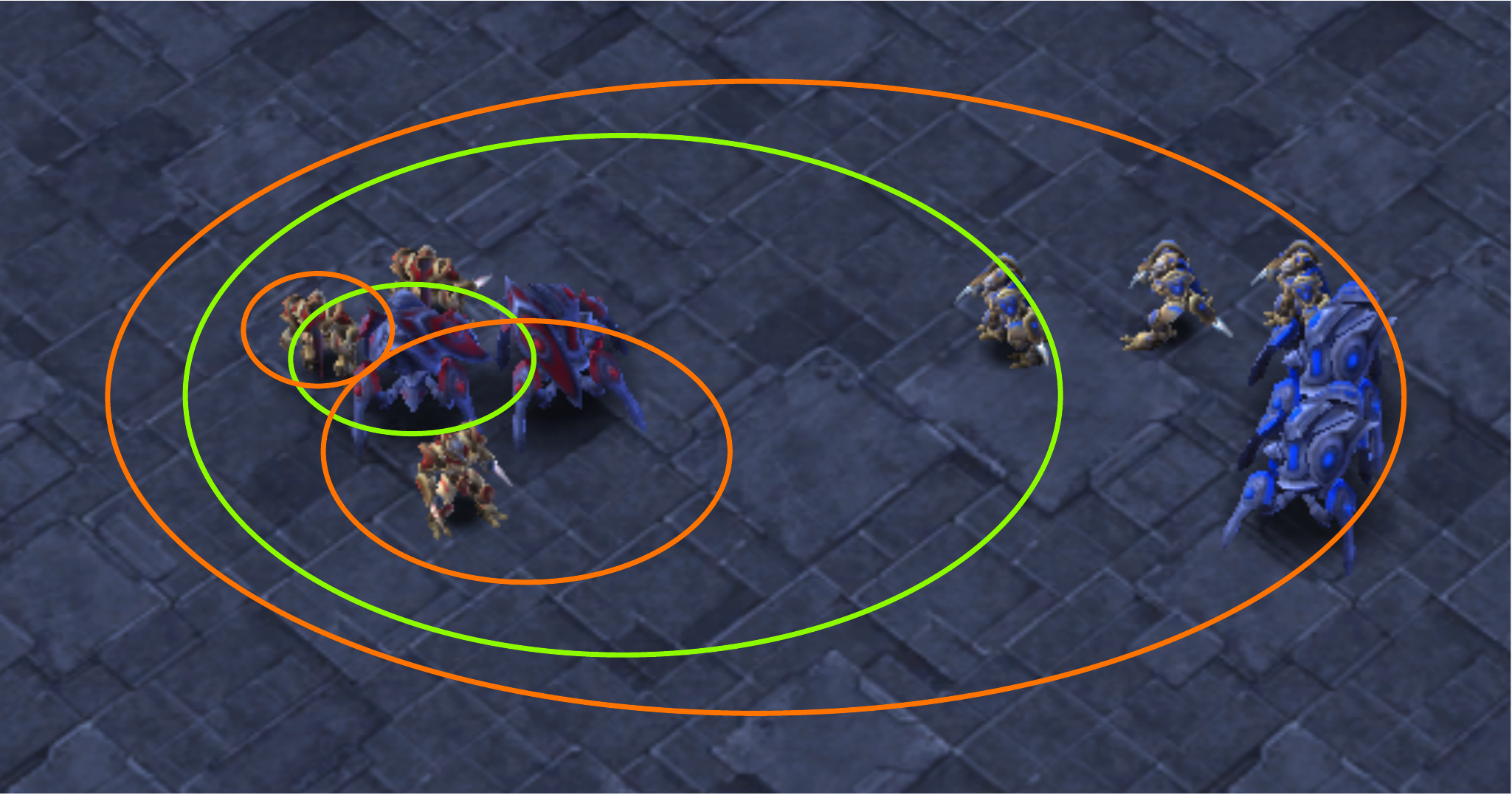}
}
\hspace{-0.5em}
\subfigure[SMAC 3s5z] {
 \label{fig:sc2-d}     
\includegraphics[width=0.226\textwidth]{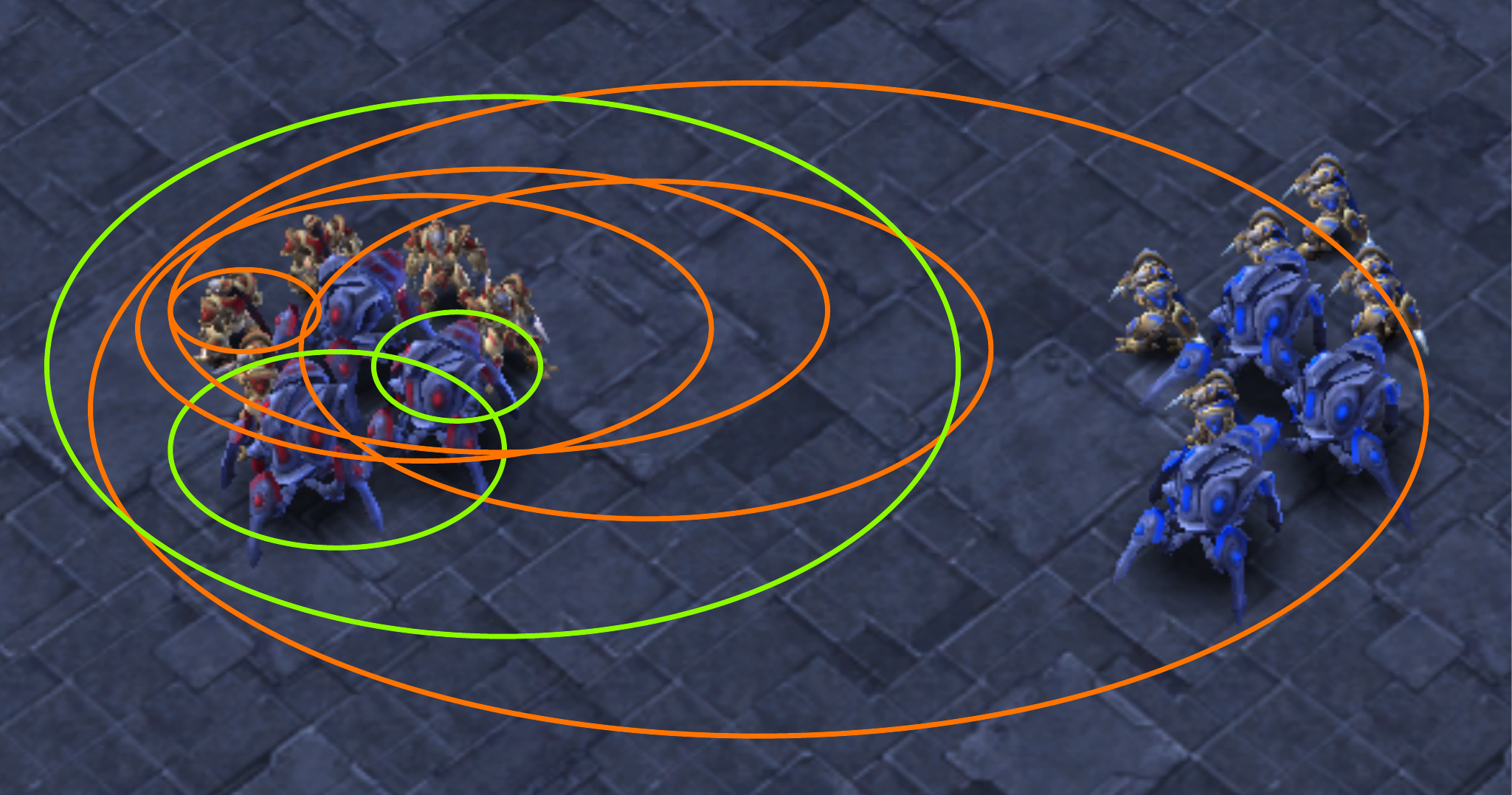}
}
\caption{Multi-agent environments.}
\label{fig:Envs}
\end{figure}

\tb{Particle Game}.
It is a two-dimensional world with continuous space and discrete time. 
We perform experiments in two scenarios, as illustrated in Fig. \ref{fig:cn} and Fig. \ref{fig:pp}.
The darker the color, the wider the observation range.

\noindent
(1) \textit{Cooperative Navigation (CN)}.
In this game, six agents cooperatively reach six unmovable landmarks and avoid collisions.
Agents have limited observation that contains only the relative positions of $k$ nearest landmarks and other agents.
Agents are assumed to have heterogeneous ability in observation, with $K_n = [0, 0, 1, 2, 4, 5]$. 
For example, the first two agents can only observe themselves, and the last agent can observe five landmarks and all the agents.
Besides, agents can observe the communication delay of the other agents.
Ideally, agents make actions based on local partial observation and messages received by other agents to cover landmarks as quickly and numerous as possible. As we introduce communication delay into this scenario, agents need to make trade-off between the gains of receiving a message and the delay in action-making.

\noindent
(2) \textit{Predator-Prey (PP)}. 
In this game, we set four slower predators chasing four faster preys, and two landmarks block the way.
Preys are assumed to do random actions, and predators are trained using DACOM or the benchmark algorithms.
Agents are rewarded based on the distance to the nearest prey among all agents, obtain a bonus by capturing prey, and punish when colliding with other agents.
Each predator can observe the relative positions and velocities of $k$ nearest predators and preys, the positions of all the landmarks, and the communication delay to the other predators. The predators are also assumed to have different observation abilities, which is $K_n = [0,1,2,3]$.
In this game, predators collaborate to catch as many preys as possible.

\tb{Traffic Control}.
Fig. \ref{fig:tc} shows an autonomous driving environment
with multiple vehicles driving through an intersection while avoiding collisions.
We set the number of autonomous vehicles to four, with a random initial position on the road.
Similar to the previous scenario, the agents obtain different observation abilities, \ie, rectangles with $K_{rec} =[5\times3, 5\times3, 5\times3, 20\times 20]$ (in meters).
The agent receives a punishment if a collision occurs and a bonus if it successfully reaches the destination in time. 
In such conditions, to obtain high rewards, agents need to move at high speed in the right direction while avoiding collisions.

\tb{StarCraftII}.
For SMAC, we consider four combat scenarios (\ie, 3m, 8m, 2s3z, and 3s5z), as shown in Fig. \ref{fig:sc2-a}-(g).
The goal of the games is to control the allies (agents) to destroy all the enemies while minimizing the total damage taken by the allies.
The observation range of the agents is set to be a circle with radius $K_r = [0.1, 5, 12, 0.2, 9, 9, 9, 20]$ respectively, which means only the enemies and the allies in the circle can be observed.
The observation radius is taken sequentially in order from $K$.
For example, the first 5 values (\ie, $K_{r} = [0.1, 5, 12, 0.2, 9]$) are selected as observation range for the five agents in 2s3z.

\tb{Communication Environment Settings}.
We generate various delays by setting different sizes of maps because it changes the mean distance of the agents and eventually affects communication delays.
To illustrate different delay environments, we define the mean delay ratio $ \varpi = \frac{\bar{d}}{\Delta t}$, where $\bar{d}$ is the mean delays of maps given randomly generated agent locations and $\Delta t$ is the time between two adjacent steps.
For example, if agents observe the environment and select actions every $\Delta t= 100\, ms$ with mean delay ratio $\varpi =10\%$, the mean communication delay between two randomly located agents is $10\, ms$.
It is worth noting that the distribution of delays varies with games even with the same $\varpi$.
For example, the particle game follows a normal distribution (see analysis in appendix) and SMAC follows other distributions (see Fig. \ref{fig:sc2}).

\subsection{Benchmark Algorithms and Hyperparameters}

We compare DACOM with MADDPG \cite{lowe2017multi}, ACML \cite{mao2020learning}, ATOC \cite{jiang2018learning}, GACML \cite{mao2020learning}, and SchedNet \cite{kim2019learning}.
The main characteristic of these algorithms is shown in the table in the appendix.
MADDPG trains the policy network for each agent independently without any communication.
ACML uses a full communication model in which the agents can obtain messages from all the other agents.
To diversify the algorithm comparison, the communication of ACML is set to be distributed. 
ACML can be designed as a centralized aggregator, but it is less effective because more latency is usually introduced by forwarding messages through a central node.
GACML and SchedNet can select agents to transmit messages as participants and have centralized aggregators.
SchedNet is set with limited bandwidth, allowing only two agents to share information.
ATOC has a gate in each agent to decide whether the agent should wait for aggregated messages from a central aggregator.
Besides, benchmark algorithms with a fixed timer  (\ie, 15\% and 35\%) are used as baselines.
Here, we define $D_c$ as the round-trip delay between agents and centralized aggregators. 
If delays follow a normal distribution, $D_c$ should have the mean value $2\lambda + \xi_n \sigma$. 
However, some baselines, such as SchedNet, aim to reduce communication with limited bandwidth, reducing delays.
Therefore, we give a discount on centralized delay ratios, which are assumed to be constant values, $D_c= [15\%, 40\%, 60\%, 80\%, 95\%]$ for the communication environment with a mean delay of [10\%, 30\%, 50\%, 70\%, 90\%], respectively.


In the experiments, we use an Adam optimizer with a learning rate of 0.005. The discount factor for reward, $\gamma$, is 0.95. For the soft update of target networks, we set $\xi = 0.01$.
We use a three-layer multilayer perceptron (MLP) with 64 units for the Encoder and four-layer MLP with 64 units to implement the TimeNet, the ActorNet, the CriticNet, and other networks in baselines, such as weight generators of SchedNet and gates in GACML.
The neural networks use ReLU as activation functions.
We initialize the parameters with random initialization.
We train our models on Intel Core i7-8700K CPUs. The capacity of the replay buffer is $10^5$, and we take a minibatch of $1024$ to update the network parameters. 
The aggregator is implemented by an attention-based unit with two attention heads. Dimensions of messages, the output of the Encoder and the aggregator, are six.

\subsection{Results}

\begin{figure}[h!]
\centering
\hspace{1.2em}
\includegraphics[width=0.416\textwidth] {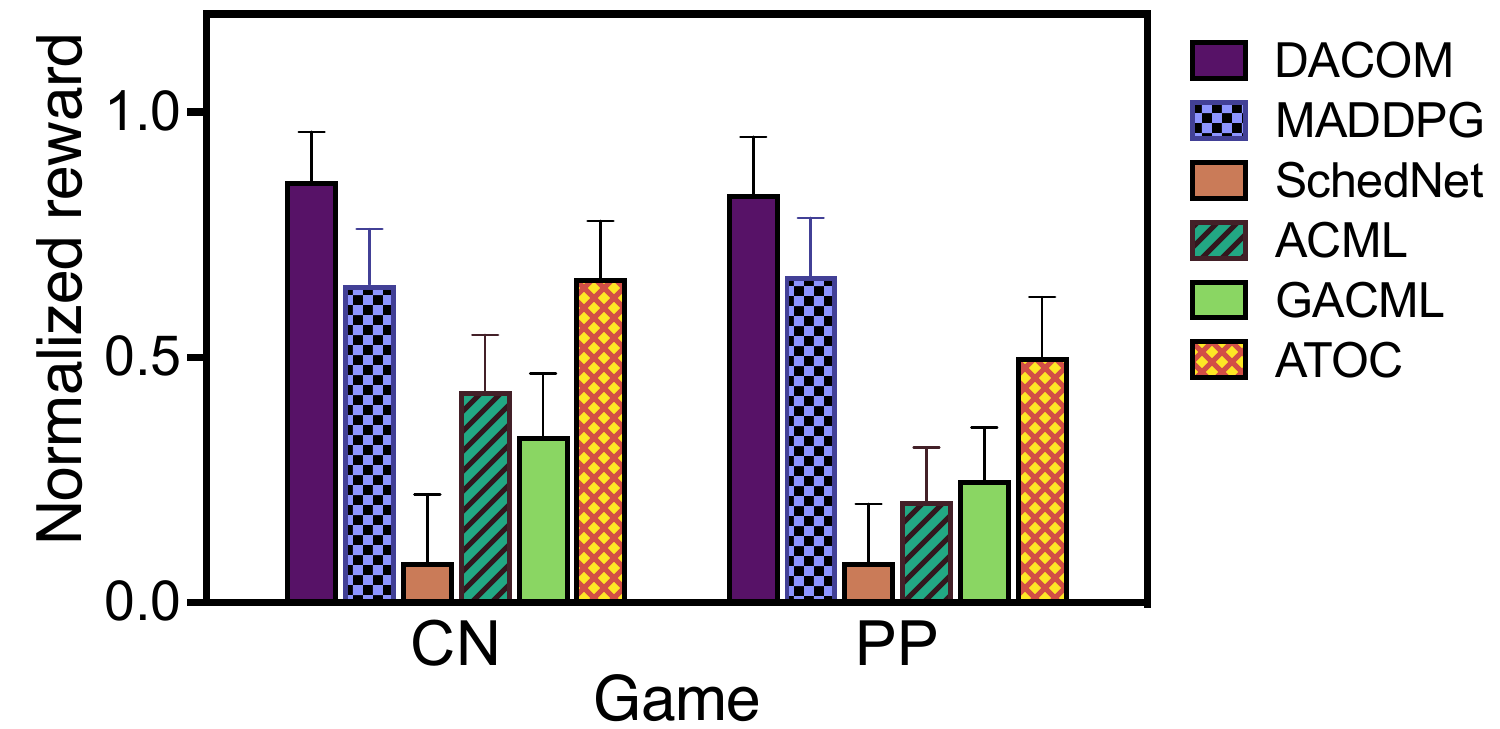}
\caption{DACOM vs. baselines in particle games under mean delays of 30\%. It shows normalized rewards with a 95\% confidence interval.}
\label{fig:Particle_Game}
\end{figure}

\tb{Particle Game}.
Fig. \ref{fig:Particle_Game} illustrates the normalized rewards against the baselines under mean delays of 30\% in CN and PP.
DACOM can achieve higher rewards than baselines.
Because agents of DACOM are delay-aware and can adaptively set waiting time to make a good trade-off between the costs of delays in action-making and gains of obtaining messages.

\begin{figure}[h!]
\centering
\hspace{-2.7em}
\includegraphics[width=0.33\textwidth] {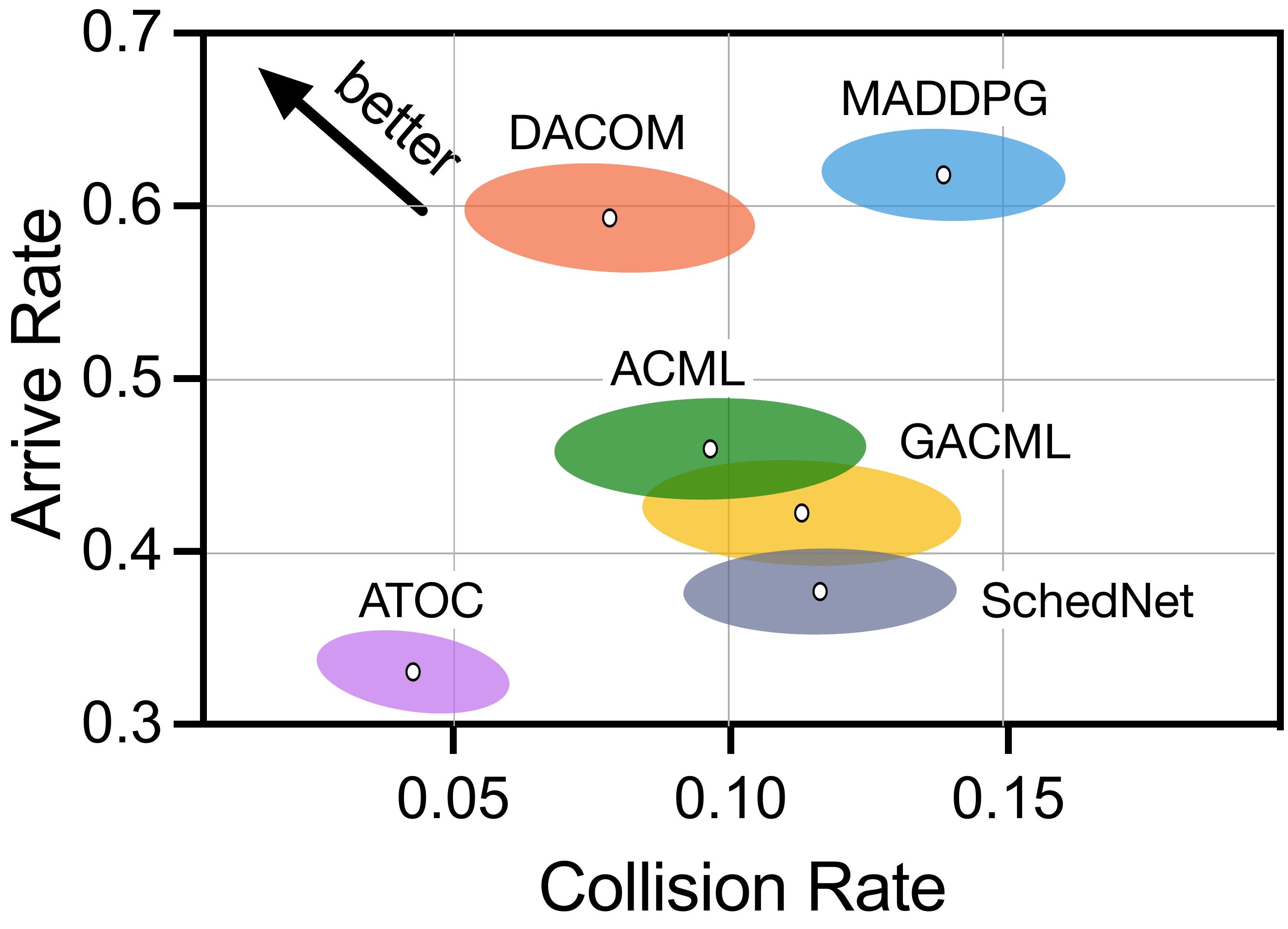}
\caption{DACOM vs. baselines in the intersection scenario under mean delays of 10\%. It shows the mean arrival rate and collision rate with a 95\% confidence interval. DACOM can achieve fewer collisions without compromising much arrival rate (even increasing some) than the baselines.}
\label{fig:highway_pop}
\end{figure}

\tb{Traffic Control}.
We illustrate the performance in intersection scenarios.
According to the reward definition, this game aims to trade-off between low collision and a high arrival rate.
Fig. \ref{fig:highway_pop} shows the collision rate and arrival rate of DACOM against baselines.
DACOM achieves a lower collision rate and a higher arrival rate compared to baselines.
Especially, ATOC achieves lower collisions than DACOM by obviously sacrificing arrival rates, which results in lower rewards (see Appendix).
On the contrary, MADDPG achieves a higher arrival rate than DACOM by greatly increasing collision, resulting in lower rewards as well.

\begin{figure*}[htp!]
\centering
\subfigure[normalized rewards with delays] {
 \label{fig:pd-a}     
\includegraphics[width=0.317\textwidth]{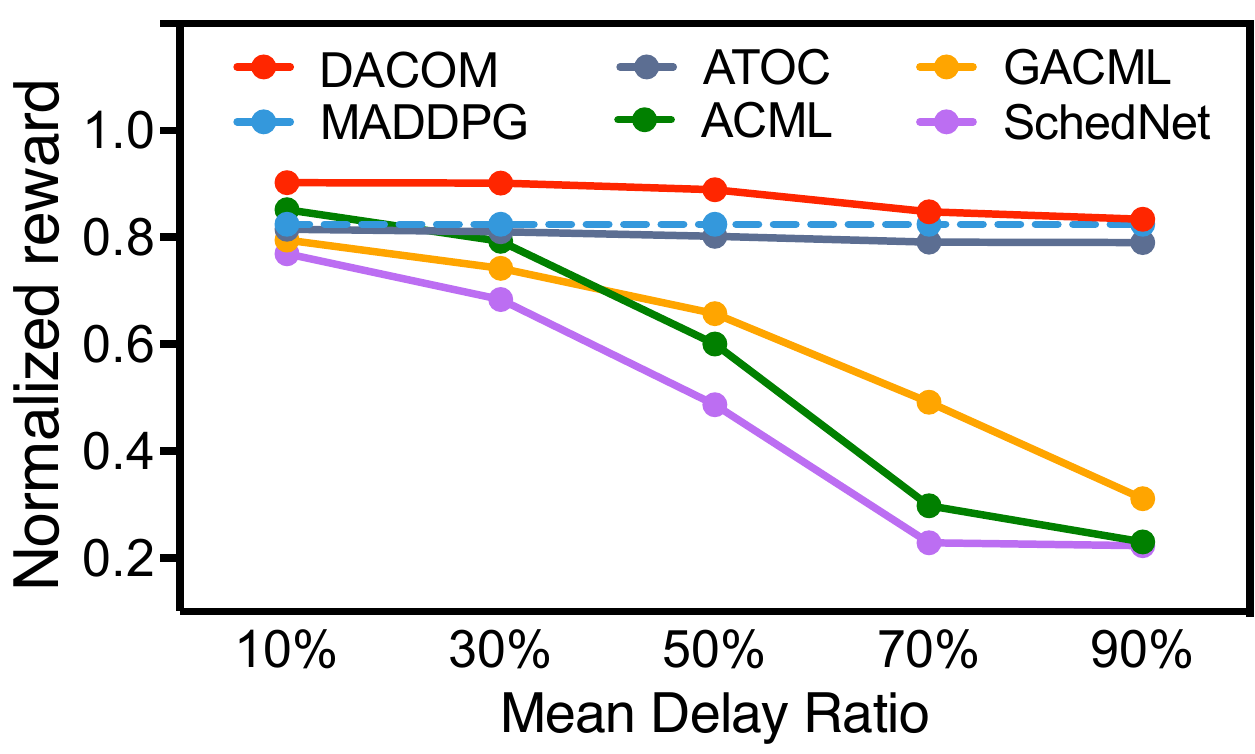}
}
\hspace{0.5em}
\subfigure[DACOM vs. fixed timers] {
 \label{fig:pd-b}     
\includegraphics[width=0.317\textwidth]{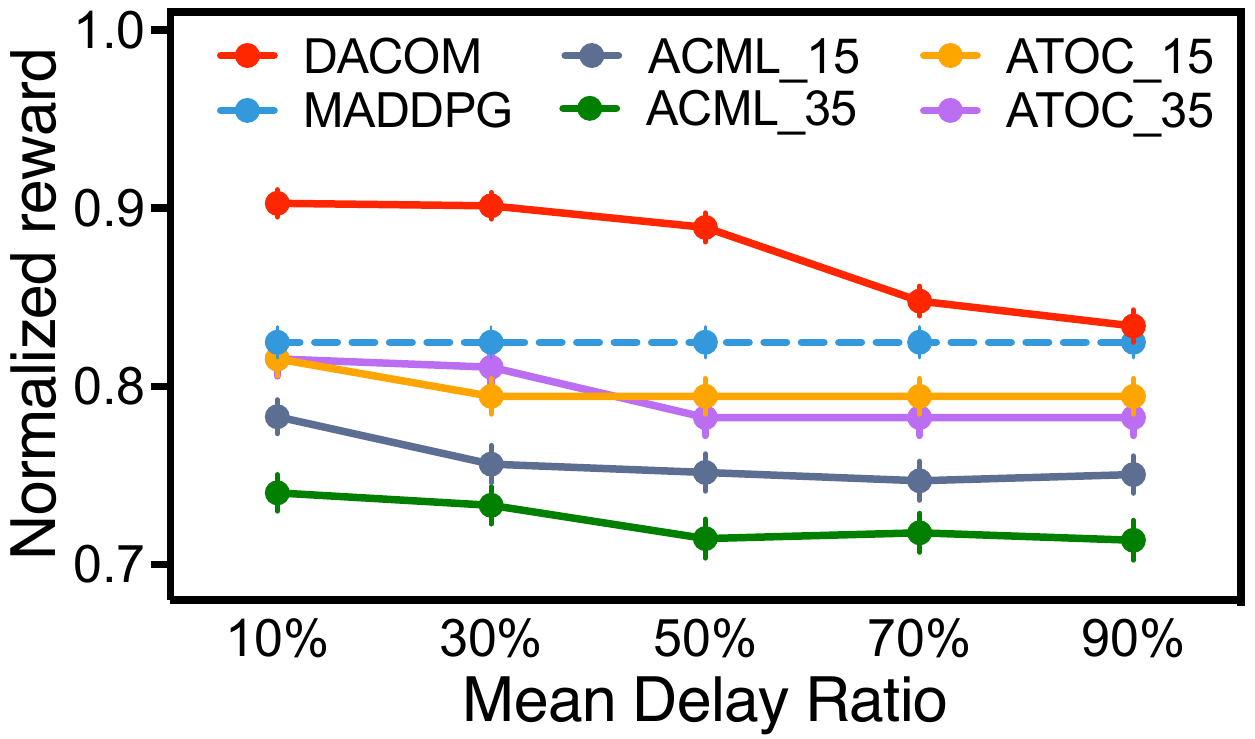}
}
\hspace{0.2em}
\subfigure[action delays with delays] {
 \label{fig:pd-c}     
\includegraphics[width=0.302\textwidth]{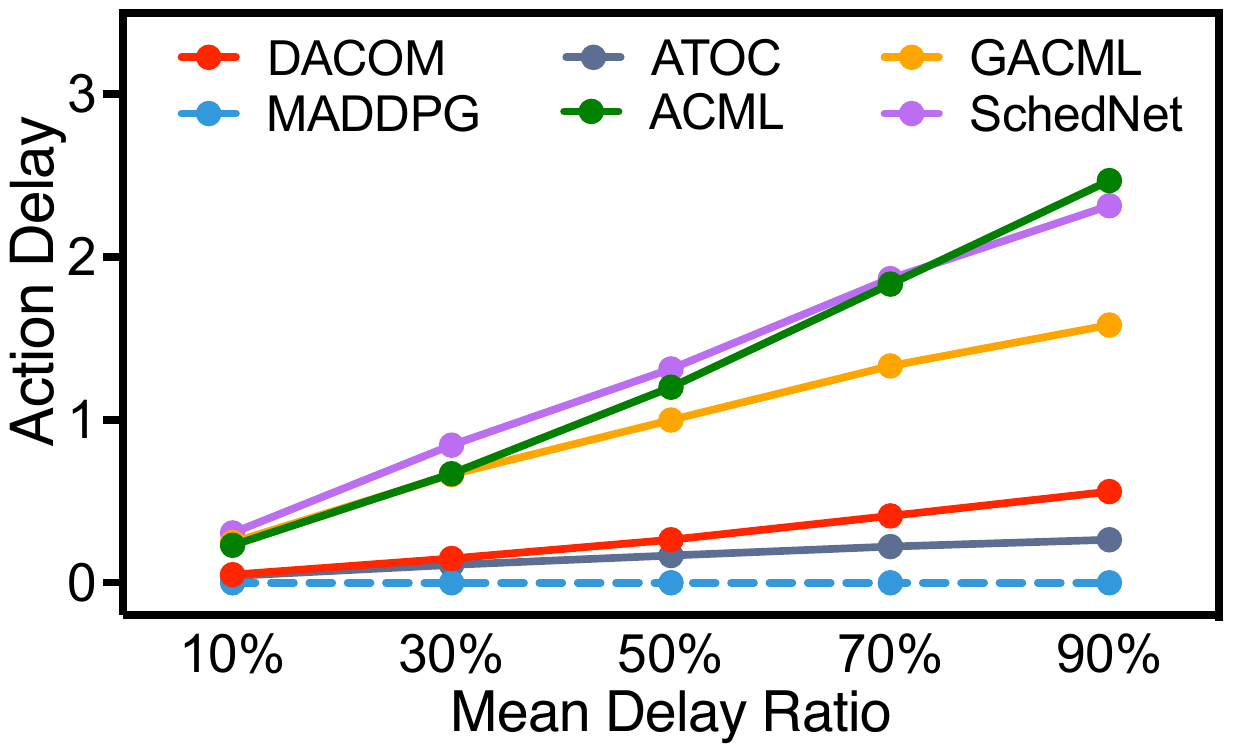}
}
\caption{DACOM vs. baselines in resilience to delays in the predator-prey game. DACOM has better resilience to delays than the baselines (including MARL with fixed timers).}
\label{fig:particle_delays}
\end{figure*}

\begin{figure}[h!]
\centering
\hspace{-1em}
\subfigure[win rate] {
 \label{fig:sc2_wr}
\includegraphics[width=0.37\textwidth]{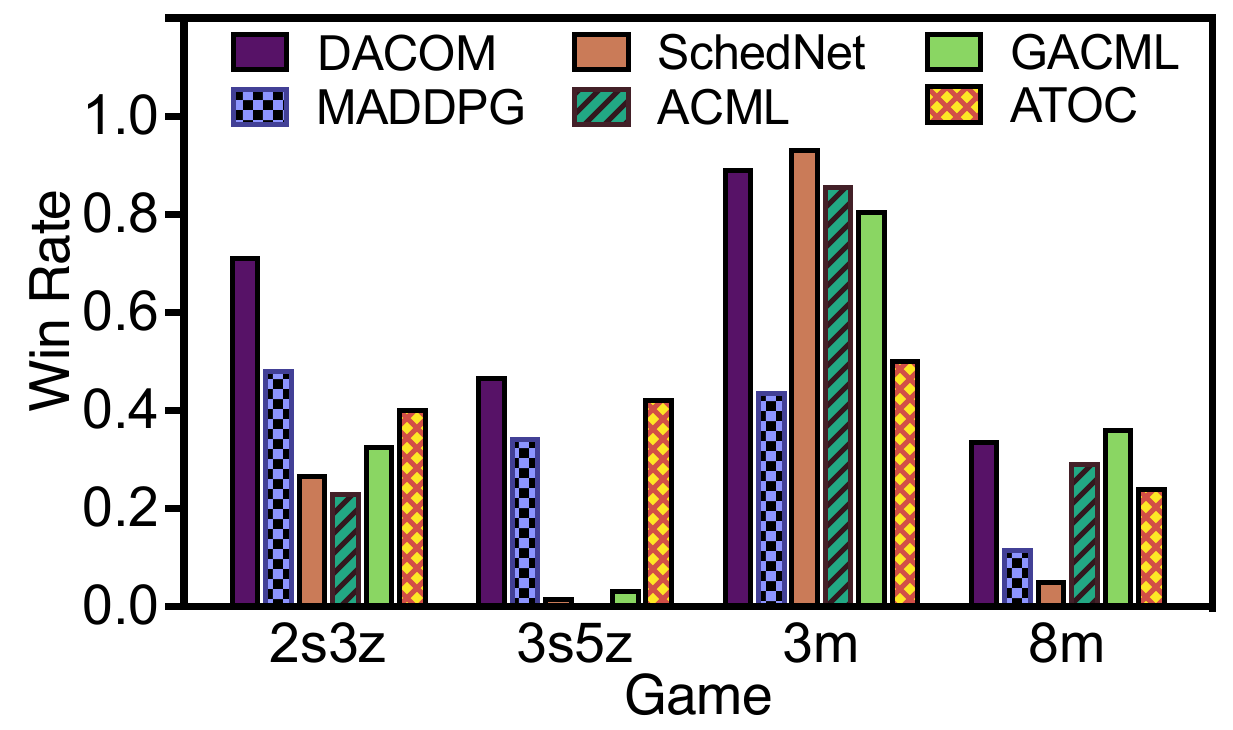}
}
\subfigure[distribution of delay] {
 \label{fig:sc2_dd} 
\includegraphics[width=0.365\textwidth]{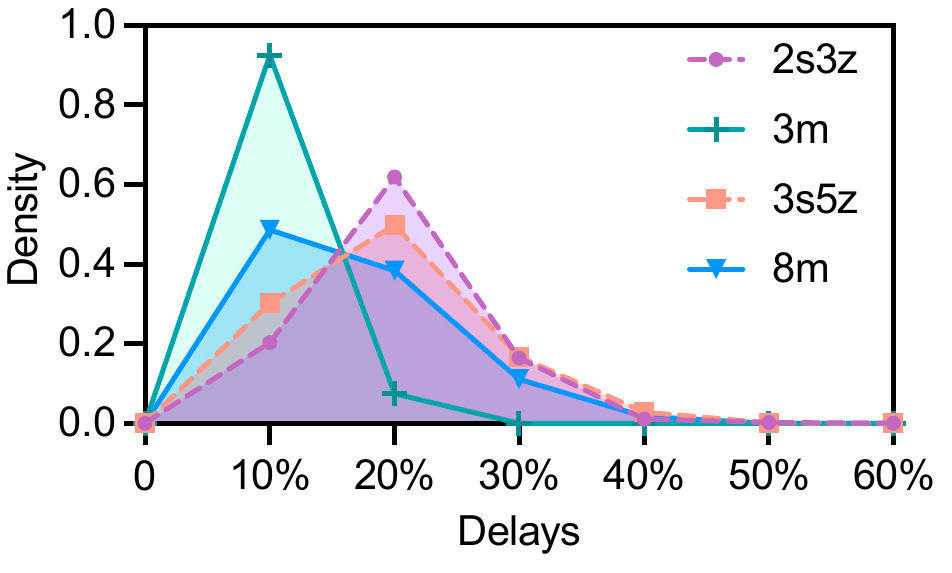}
}
\caption{DACOM vs. baselines in SMAC under mean delays of 30\%. DACOM has better performance for 2s3z and 3s5z, and less improvement on 3m and 8m due to different delay distributions.}
\label{fig:sc2}
\end{figure}

\tb{StarCraft II}.
We apply our method and baselines to SMAC.
We select four scenarios (\ie, 2s3z, 3s5z, 3m, and 8m) for performance validation under 30\% mean delay settings.
The results show that DACOM generally has a good win rate (see Fig. \ref{fig:sc2_wr}).
Unlike the previous two games, SMAC is more challenging due to different types of agents (\eg, stalkers and zealots) and the need for group cooperation.
For example, in the 3m and 8m, agents are subject to aggregated attacks and tend to gather to resist, which brings lower communication delay due to the proximity than the other games even with the same set of communication channels. (see Fig. \ref{fig:sc2_dd}\footnote{In SMAC, one timestep of MARL is ten game steps, so the successive delays are discretized into the number of game steps.}).
Therefore, DACOM has limited advantages in 3m and 8m because most delays are less than a game step whose effects can be ignored.



\tb{Why is DACOM robust against delays?}
We evaluate DACOM and the baselines in five different delay settings using PP as an example (see Fig. \ref{fig:particle_delays}) and similar results can be obtained from the other games.
The results reveal that DACOM has the highest rewards than the baselines and fewer action delays than most of the baselines.
When the mean delay ratio is greater than 70\%, DACOM has similar rewards to MADDPG, which means the communication is nearly useless without any gains.
The other MARL approaches work relatively well in scenarios with low delays, but the performance is significantly dropped as delays grow because they are designed for a delay-free environment and cannot keep good performance in high delays.
Furthermore, the ACML and ATOC with a fixed timer (15\% and 35\% for ACML, 15\%, and 35\% for ATOC) are used as baselines (see Fig. \ref{fig:pd-b}).
We can see that the predefined timer can
prevent reward reducing significantly with communication delay increases by preventing agents from waiting too long, comparing Fig. \ref{fig:pd-b} and Fig. \ref{fig:pd-b}.
However, ACML's rewards with the fixed timer are lower than those without a fixed timer when the delay is approximately 10\%. 
That is, the fixed-timer also limits communication gains and cannot adapt to changing environments.




\section{Conclusion}
\label{con}
In this paper, we demonstrate that ignoring communication delay results in performance degradation, which is evident in delay-sensitive tasks and high-delay environments.
To improve agent cooperation against the influence of communication delay, we have proposed DACOM, a delay-aware communication model for MARL.
DACOM is an adaptive delay-aware agent communication model in which agents learn to schedule how long to wait for messages from other agents.
Unlike the existing methods, DACOM can reduce the uncertainty of cooperation outcomes in dynamic communication networks and improve agent cooperation by adapting communication to network states. 
Empirically, DACOM outperforms existing methods in various cooperative multi-agent environments, especially obvious in delay-sensitive tasks.

\noindent
\textbf{Limitations}.
DACOM does not perform better than existing schemes in all scenarios, such as delay-insensitive tasks, scenarios with low delay ($<$ 10\%) or very high delay ($>$ 70\%), and low gains of communication (\eg, all agents are well-informed).
Besides, our target is to improve agent cooperation through communication, and therefore the delays, except for communication delays, are not discussed in this paper.
Finally, personal information leakage may happen when utilizing the communication between agents. We do not offer additional protection for this case because we assumed agents are fully cooperative with no privacy concerns.


\section{Acknowledgments}

This work has been partly funded by EU H2020 COSAFE project (Grant No. 824019), Horizon Europe CODECO project (Grant No. 101092696), the National Natural Science Foundation of China (No. 62002028), the Alexander von Humboldt Foundation.


\bibliography{ref}

\clearpage

\end{document}